




\magnification=1200
\overfullrule=0pt
\parskip=10pt

\def\.{\!\cdot\!}
\def\({\left(}
\def\){\right)}
\def\[{\left[}
\def\]{\right]}
\def\h{{1\over 2}}
\def\b{$$\eqalignno}
\def\p{\partial}
\def\bk#1{\langle#1\rangle}
\def\Tr{{\rm Tr}}
\def\tr{{\rm tr}}
\def\bbl{\left\{}
\def\bbr{\right\}}
\pageno=0
\rightline{McGill/92-53}
\rightline{hep-ph/9212296}
\rightline{(December, 1992)}
\bigskip\bigskip\bigskip
\centerline{\bf MULTILOOP STRING-LIKE FORMULAS FOR QED}
\bigskip\bigskip\bigskip
\centerline{C.S. Lam$^*$}
\bigskip
\centerline{Department of Physics, McGill University, 3600 University St.}
\medskip
\centerline{Montreal, P.Q., Canada H3A 2T8}
\bigskip\bigskip\bigskip\bigskip
\bigskip\bigskip
\centerline{\bf Abstract}

\bigskip\bigskip\bigskip
Multiloop gauge-theory amplitudes written in the Feynman-parameter
representation are poised to take  advantage of  two important
developments of the last decade: the spinor-helicity technique
and the superstring reorganization. The former
has been considered in a previous article; the latter will be elaborated
in this paper.
We show here how to write multiloop string-like formulas in the
Feynman-parameter representation for any process in QED, including those
involving other non-electromagnetic interactions.
The general connection between the Feynman-parameter approach and
the superstring/first-quantized approach is discussed.
In the special case of a one-loop multi-photon amplitude,
these formulas reduce to the ones obtained by the superstring and
the first quantized methods. The string-like formulas exhibits a simple
gauge structure which makes the Ward-Takahashi
identity apparent, and  enables  the
integration-by-parts technique of Bern and Kosower
to be applied, so that gauge-invariant parts can be extracted
diagram-by-diagram
with the seagull vertex neglected.

\vfill
\leftline{*\ email address: Lam@physics.mcgill.ca}
\vfill\eject

\centerline{\bf 1. Introduction}

Feynman diagram calculations for gauge theories are always complicated
because of the presence of
spin and gauge dependence. Spin brings in derivative coupling or Dirac
algebra; gauge dependence introduces many gauge non-invariant terms
which  will be cancelled  at the end.
 For non-abelian theories
such as QCD the situation is even worse, for there is the additional
complication of color algebra, as well as the presence of  three- and
four-gluon vertices, and  ghosts.

Two techniques have been developed in the last decade to simplify
gauge theory calculations for tree and  one-loop diagrams. The first is the
{\it spinor-helicity technique} [1--21], which makes use of the fact that most
fermions can be treated as massless at high energies. Being massless,
chirality is conserved, a fact which
 can be exploited effectively to reduce
the number of terms in a significant manner. Even photons and gluons benefit
from this conservation because a spin-one index can be written as a pair of
dotted and undotted spinor indices. Furthermore, one may make use of gauge
freedom to choose the  polarization vectors wisely to
render many more terms zero.
Polarized amplitudes benefit enormously from this
technique but even unpolarized cross sections for which this technique
was first invented are
greatly simplified.
This technique, however, is applicable only when
all the momenta in the problem are linear combinations of massless momenta.
This is so for tree diagrams, but not true  for loop diagrams where
off-shell loop-momenta are present
to ruin chirality conservation.
Consequently most of the applications of this technique has been confined
to tree amplitudes [1--21].

This problem can be circumvented if the off-shell loop-momenta are
integrated out before the spinor-helicity technique is applied.
One way of doing so is to make use of the connection between
a gauge theory and a superstring theory in the infinite tension
limit [22--24]. There is a known formula for a one-loop superstring
amplitude
in which integrations are over the Koba-Niesen variables with
 internal momenta  absent. This formula for
a one-loop multi-gluon (or multi-photon) amplitude
can also be derived from a first-quantized particle theory interacting with
a background gauge field [25], where the particle in the theory is taken to be
the one running around the loop. The formula obtained
 in the first-quantized theory is identical
to the superstring formula
presumably because the particle theory used in this approach is also
reparametrization invariant.

Unfortunately, the generalization of either of these two approaches
 beyond one loop, and to all possible scattering amplitudes, have
not yet been worked out.

A different method to avoid the off-shell momenta is to introduce
Feynman parameters to integrate them out [26].
Spinor helicity technique is then applicable in the resulting
Feynman-parameter representation, in any process and to any
number of loops [27].

The second technique developed in the last decade is independent of
the spinor-helicity technique, but works very well in conjunction with it.
It can perhaps be labelled
as {\it string reorganization} [23].
For tree and certain one-loop amplitudes when string
formulas are available, one finds that the individual terms
of the string formula in the
infinite-tension limit
 are not identical to
the terms obtained from the standard Feynman rules,
although the sum of all the terms are necessarily the same for both.
Moreover, as a rule, the terms in the string formula are organized in a
much neater and a much simpler way, and this beautiful reorganization
can greatly simplify calculations.
 Why the string is so
clever is not completely clear; it may have to do with the fact that it
treats spacetime and color on an equal footing.
I shall refer to
field-theoretic formulas reorganized in the string way as {\it string-like
formulas}.

Because of its neatness and simplicity, and potential simplifications
achieved in actual calculations,
it would be desirable to develop string-like formulas for
all processes in any number of loops. It is not a priori clear that
this can be done, and if so in what form the string-like formulas
would take, because we only know them for some amplitudes and only
to one loop order. Fortunately it turns out that in many cases the string-like
formula in one loop is sufficient of a guide for obtaining
 string reorganization
in many loops. For example,
in tree and one loop QCD, string [4, 15--17]
reorganizes the usual color factors into
the Chan-Paton factors [28], which have the advantage of having
the corresponding subamplitudes gauge invariant and
cyclic-permutation invariant. This reorganization can be generalized
to all loops and all amplitudes [27].

Background gauge emerges naturally from the one-loop string
calculations of QCD [23] and the first quantized approach [25].
Its use simplifies the calculations considerably in any number of loops
as well,
especially when a large number of external gluon or photon
 lines are present [27]. This again illustrates the superiority of
string reorganization.

There is yet another important feature in the one-loop
string-like formula  which is the main subject of this paper.
I shall refer to this feature as the {\it gauge characteristic}, because
it makes explicit the gauge-transformation property
and the Ward-Takahashi identity in the
Feynman-parameter space. To see its significance take
a {\it momentum-space}
diagram in spinor QED with some external photon lines.
A gauge transformation of the polarization vector produces two
additional terms given by
shrinking  one of the two spinor propagators
this external photon is attached to. It is this local feature which the
 Ward-Takahashi identity is depended on. In {\it Feynman-parameter space},
this local feature is
lost, because rules governing Feynman-parameter amplitudes tend
to involve the diagram globally as a whole [26]. Nevertheless,
when the amplitude is reorganized into a string-like formula,
this local feature reappears once again.
 The details will be
discussed in Sec.~3, but what happens is that in this string-like
form, the additional terms arising from
the gauge transformation are proportional to
derivatives of the integrand wrt some  Feynman parameters $\alpha _r$. The
surface terms obtained by integrating these derivatives  correspond
to diagrams with $\alpha _r$ set equal to zero,
which is equivalent to shrinking and short-circuiting the corresponding
propagators.
In this way the essential local feature of the momentum-space
leading to the Ward-Takahashi
identity is restored. This is the gauge characteristic alluded to before.

This gauge characteristic of the string-like formula has been put to good
use  in the known one-loop string-like formulas [22--25].
To explain it let us define a diagram
to be a {\it standard diagram} when none of the {\it external}
gauge-particle lines are connected to the diagram either through
a seagull vertex (in scalar electrodynamics)
or through a four-gluon vertex (in QCD). All other diagrams will be called
{\it seagull diagrams}. Because of the presence of this
gauge characteristic in the {\it standard diagrams}, it is possible to perform
integration by parts (IBP) in the Feynman parameters. At least in
simple cases, the gauge-dependent surface terms
together add up to cancel contributions from the seagull diagrams.
Consequently we never have to worry about the surface terms nor the
seagull diagrams. This greatly simplifies the calculation because
there are many many seagull diagrams, and because in this way
the gauge-invaraiant parts can be extracted
{\it diagram by diagram} after discarding the surface terms from suitable
IBP's.

This   gauge clarity of the string-like formula and the resulting
simplification from the IBP technique makes it important to
ask whether string-like formulas can be written
for all loops and all processes. We shall leave non-abelian gauge
theories aside for future considerations. The purpose of this paper is to
show that string-like formulas with the gauge characteristic

can be written for standard diagrams in
QED, in the Feynman-parameter representation to any number of loops and for
all
processes, provided a technical restriction to be discussed in Sec.~3 is
obeyed.

 In order to dervive such multiloop formulas,
considerable knowledge of the details of the Feynman-parameter representation
and the electric circuit analogy [26, 29--32] is needed. We therefore start
out
in Sec.~2 with a discussion of these topics. Some but not all of the formulas
appearing there are already contained in [26], but the important
and subtle roles of {\it external vertices} and of
{\it level dependence} were not sufficiently
discussed before. Currents in an electric circuit are the important objects to
be considered in momentum-space
amplitudes, and voltages are the primary quantities to
be dealt with in the configuration-space amplitudes. Voltage at a vertex
corresponds to the position of that point in the configuration space.
Translational invariance
in configuration space correponds to the impossibility of determining
the absolute (rather than the relative) voltage levels, which in turn is the
cause of the level-dependence problem.
Though there is no direct relationship, it is
however interesting to note that
in many ways level-dependence in an electric circuit is analogous to
gauge-dependence in a gauge theory.

Together with the electric circuit analogy of a Feynman diagram it is also
known that there is a particle
interpretation of the same, which as we shall see
 comes very close to being a multiloop generalization of
the string approach of Bern and Kosower [22]
and the first quantized
approach of Strassler
[25]. This will also be discussed in Sec.~2.

One feature of the string-like formula is that it treats external vertices
somewhat differently than the internal vertices. This can be seen for example
in the appearance of the background gauge in QCD [23, 25].
We shall see in Sec.~2 that to some extent this is a general feature,
in that the topological structure of Feynman diagrams is such that
certain relations which hold for external vertices are no longer true
for internal vertices.

The multiloop string-like formula for
scalar QED with the gauge characteristic  will be derived in Sec.~3.
A similar formula for spinor
QED will be discussed in Sec.~4.
An explicit two-loop example for a photon-meson Compton scattering diagram is
given in Sec.~5 to illustrate some of these results.

\bigskip

\centerline{\bf 2. Feynman-parameter Representation}

Any $\ell$-loop scattering amplitude with $N$ internal and $n$ external
lines is given by a momentum-space integral
$$M(p)=\left[ {-i\over
(2\pi)^4}\right]^\l\int\prod_{a=1}^\ell(d^4k_a){S_0(q)\over
\prod_{r=1}^N(-q_r^2+m_r^2-i\epsilon)}\ ,\eqno(1)$$
where the internal momenta $q_r$ are understood to be linear combinations of
the external {\it outgoing} momenta $p_i$ and the loop momenta $k_a$.
An external momentum $p_i$ will be assigned to {\it every}
vertex $i$; this  allows certain algebraic manipulations that would otherwise
be impossible.
At the end of the calculation all
artificially added $p_i$  will
be set equal to zero.

Vertex factors as well as numerators of spinning propagators are incorporated
into the {\it `primitive spin factor'} $S_0(q)$.
We shall work in four dimensions but
it is just as easy to develop formulas for an arbitrary dimension.
By introducing the Feynman parameters $\alpha_r$, the loop momenta
can be integrated out and a Feynman-parameter representation
of the amplitude can be obtained [26]:
$$\eqalignno{
M(p)&={i^N\over (- 16\pi^2)^\l}
\int_0^\infty (\prod_{r=1}^Nd\alpha_r)  \Delta(\alpha )^{-2}
\exp[-iD(\alpha,p)]S(J)\ ,&(2)\cr
D(\alpha ,p)&=\sum_{r=1}^N \alpha _rm_r^2-P(\alpha ,p)\ ,&(3)\cr
P(\alpha,p)&=\sum_{r=1}^N\alpha_rJ_r^2
\ .&(4)}$$
Eq.~(2) is written in the Schwinger proper-time formalism, or
the `Nambu representation'. One can recover from it the `Chisholm
representation' quoted in [27] by making the substitution
$\alpha_r=\bar \alpha_r\lambda$, with $\sum_r\bar \alpha_r=1$, and carrying out
the integration over $\lambda$.

To understand and to describe the quantities $J,P,\Delta$, and $S$ appearing in
(2), it is useful to know that
a Feynman diagram can be thought of as a passive electric circuit [26, 29--32],
in
which $\alpha_r$ takes on the role of resistance and $p_i$
become the currents flowing out of the circuit.
The quantity $J_r$ is then the current
flowing through the $r$th internal line, and $P$ is
the power consumed by the circuit.
Explicit formulas for these quantities are available [26],
$$\eqalignno{
\Delta(\alpha )&=\sum_{T_1}(\prod^\ell \alpha)\ ,&(5)\cr
P(\alpha ,p)&=\Delta (\alpha )^{-1}\sum_{T_2}(\prod^{\ell+1}\alpha )
(\sum p)^2\ ,&(6)\cr
J_r(\alpha,p)&=\pm\Delta (\alpha )^{-1}\sum_{T_2(r)}\alpha _r^{-1}
(\prod^{\ell+1}\alpha )(\sum p)\ ,&(7)}$$
and they mean the following.
An $\ell$-loop diagram can be made into a connected tree diagram (a `1-tree')
by cutting $\ell$ lines, and into a diagram with two disjoint trees
(a `2-tree') by cutting $\ell+1$ lines. $\Delta(\alpha)$ is given by
the sum over the set $T_1$ of all 1-trees so obtained, with the summand
consisting of
the product of the $\alpha$'s of the cut lines. $P(\alpha,p)$
is given by the sum over the set  $T_2$ of all 2-trees so obtained, with the
summand being
the product of the $\alpha$'s of the cut lines, times the square
of the sum of all the external momenta $p_i$ attached to one of these two
trees. Finally, let $T_2(r)$ be the set of all 2-trees in which line
$r$ is cut, and such that when the line $r$ is inserted back
a 1-tree results. Then
$J_r(\alpha,p)$ is given by the sum of all 2-trees $T_2(r)$
 with the summand equal to
the product $\alpha$'s of all the cut lines except the $r$th, times the
sum of all the external momenta $p_i$ attached to one of these two trees.
The sign $\pm$ in (7) is determined by the orientation of $J_r$.

We proceed now to describe the {\it `modified spin factor'} $S(J)$ in (2). It
is made up of the sum of several terms,
$$S(J)=\sum_{k\ge 0}S_k(J)\ ,\eqno(8)$$
of which the first, $S_0(J)$, is just the numerator factor $S_0(q)$ in (1)
with $q$ replaced by $J$. The other terms
$S_k$ are obtained from $S_0$ by contracting $k$ pairs of
$J$'s in all possible ways according to the  rule,
$$J_r^\mu J_s^\nu\to -{i\over 2}g^{\mu\nu}H_{rs}(\alpha)\ ,\eqno(9)$$
and summing over all the contracted results. The formula for $H_{rs}$ is
$$\eqalignno{
H_{rr}(\alpha)&=-\Delta(\alpha)^{-1}\partial \Delta (\alpha )/\partial
\alpha _r\ ,&\cr
H_{rs}(\alpha)&=\pm\Delta (\alpha )^{-1}\sum_{T_2(rs)}(\alpha _r \alpha_s
)^{-1}
(\prod_{\ell+1}\alpha )\ ,\quad(r\not= s)\ ,&(10)}$$
where $T_2(rs)$ is the set of all 2-trees with both lines $r$ and $s$
cut, and such that when either line $r$ or line $s$ is inserted back
 a 1-tree results.
The product of $\alpha$'s in (10) are over all the cut
lines except the $r$th and the $s$th. If  the lines $r$ and $s$
both point to the same tree, then the sign in (10) is $+1$. If
they point to different trees, then the sign is $-1$.

For practical calculations, eq.~(2) may not be the best to use
because the contractions leading up to $S_k(J)$ are relatively complicated
to compute. It may often be simpler to use the alternate formula [26]
\b{
M(p)&=
\int_0^\infty (\prod_{r=1}^Nd\alpha_r)  \Delta(\alpha )^{-2}
T\(-{i\over 2}{\p\over\p p}A\beta\) \exp[-iD(\alpha,p)]\ ,&(11)}$$
where the function $T(q)$ is obtained from the function $S_0(q)$ by the formula
$$T(q)=\sum_{k\ge 0}T_k(q)\ ,\eqno(12)$$
with $T_0(q)=S_0(q)$, and other $T_k(q)$ obtained from $T_0(q)$
by contracting $k$ pairs of $q$'s according to the rule
$$q_r^\mu q_s^\nu\to {i\over 2}\beta_r\delta_{rs}g^{\mu\nu}\ ,\eqno(13)$$
and summing over all possible contractions. The quantity $\beta_r$ is the
conductance of the line $r$:
$$\beta_r=(\alpha_r)^{-1}\ .\eqno(14)$$

The contraction rule (13) is far simpler than the contraction rule (9).
For example, if each $q_r$ does not appear more than once in $S_0(q)$,
then no contraction according to (13) is possible because of the $\delta_{rs}$
factor, so $T(q)=S_0(q)$, whereas this is not the case according to (9)
and $S(J)\not=S_0(J)$.

In (11), the argument $q_r$ in $T(q)$ is replaced by the operator
$$d_r\equiv-{i\over 2}\sum_i{\p\over\p p_i}A_{ir}\beta_r\ ,\eqno(15)$$
where $A_{ir}$ is the {\it incidence matrix} of the graph, defined to be
$+1$ if line $r$ points into the vertex $i$, $-1$ if the line points
out of the vertex, and 0 if line $r$ is not connected to the vertex $i$.
Momentum conservation at each vertex can be expressed with the help of
this matrix to be
$$p_i=\sum_rA_{ir}J_r\ .\eqno(16)$$
If line $r$ points from vertex $i$ to vertex $j$, then we shall also write
it as $r=(ij)$. In that case
(15) simply says that each $q_{(ij)}$
should be replaced by $(-i/2)\beta_{(ij)}[\p/\p p_i-\p/\p p_j]$.

Let us now look deeper into the various circuit quantities and their
relationships. Some of the relations described below already appeared in [26],
but the external-vertex relations and the level-dependent relations have not.
It is important to understand these formulas, especially
the subtle role of level-dependence, for they will be needed in deriving the
multiloop string-like formulas in Sec.~3.

Let $V_i$ be the voltage at vertex $i$, and
$$v_r=v_{(ij)}=V_i-V_j=-\sum_iA_{ir}V_i\eqno(17)$$
be the voltage drop across the resistor $r$. The current flowing
through that line is
$$J_r=\beta_rv_r\ .\eqno(18)$$
Combining (16)--(18), one gets
$$p=-(A\beta A^t)V\equiv -YV\ ,\eqno(19)$$
where $p,V$ are $n$-dimensional vectors with components $p_i$ and
$V_i$ respectively, $A$ is the $n\times N$ matrix with matrix elements
$A_{ir}$, and $\beta=\alpha^{-1}$ is a diagonal $N\times N$ matrix with
diagonal matrix elements $\beta_r=\alpha_r^{-1}$. The power consumed
by the network is then
$$P=-V\.p=VYV\ .\eqno(20)$$

The absolute level of the voltages $V_i$ are of course never determined;
they can all be shifted by a common constant without changing any physical
attribute. This is reflected
in (19) by the fact that
$$\sum_jY_{ij}=0\ ,\eqno(21)$$
 which follows algebraically from
$\sum_iA_{ir}=0$. Being a symmetric matrix, we must also have
$\sum_iY_{ij}=0$, and in (19) this simply expresses conservation of the
external
currents. The matrix $Y$ is singular, so (19) cannot be inverted to obtain
$V$ as a function of $p$, in agreement with the fact that
the absolute level of $V$ cannot be determined. This {\it level-dependence}
is analogous to a gauge dependence. It is unphysical, it complicates
matter, but often we have no choice but to fix a {\it level scheme} (analogous
to fixing a gauge) to carry out explicit calculations. For example,
we must fix a level scheme before the inversion of (19) can be carried out.

In order to invert (19), let us fix a level scheme by choosing $V_n=0$.
We shall use a subscript $0$ to denote the remaining $(n-1)$-dimensional
quantities. Then (19) can be inverted to give
$$V_0=-Y_0^{-1}p_0\equiv -Z_0p_0\ .\eqno(22)$$
Incorporating $V_n=0$,
we can enlarge this $(n-1)$-dimensional relationship
into $n$-dimensional relationship
\b{
V&=-Z'p\ ,&(23)\cr
Z'&=\pmatrix{Z_0&0\cr 0&0\cr}\ .&(24)}$$
We shall refer to this level scheme as the {\it primitive level scheme}.
The chief advantage of this scheme is that an explicit formula for
$Z'$ is available, via eqs.~(24), (22), and (19).

The impedance matrix $Z'$ is a level-dependent quantity. Because of
$p$-conservation, a change
$$Z'_{ij}\to Z_{ij}=Z'_{ij}+a_i+a_j\eqno(25)$$
for any $a_i$ simply changes the overall level of $V_i$ by an amount
$\sum_ja_jp_j$ but it
does not change the physical attributes of the network.
This level change is analogous to a gauge transformation.
Physically measurable quantities such
as the current $J_r$ and the power $P$ must be level-independent.
In fact, from (17)--(25),
\b{J&=-\beta A^tV=\beta A^tZ'p=\beta A^tZp\ ,&(26)\cr
P&=-Vp=pZ'p=pZp=J\alpha J\ ,&(27)}$$
and their level-independence follows easily from
$$\sum_ip_i=\sum_iA_{ir}=0\ .\eqno(28)$$

The variation of these level-independent quantities wrt a change of $\alpha$
is most easily calculated in the $V_n=0$ level scheme. Using (19) and
(22)--(24), one gets
\b{
{\p P\over\p \alpha_s}&=p{\p Z'\over\p \alpha_s}p=
(pZ'A\beta)_s(\beta A^tZ'p)_s=J_s^2\ ,&(29)\cr
{\p J_r\over\p \alpha_s}&=-\beta_s\delta_{rs}J_s+[\beta A^tZ'A\beta]_{rs}
(\beta A^tZ'p)_s=H_{rs}J_s\ ,&(30)}$$
in which we have used the definition of $H_{rs}(\alpha)$:
\b{
H(\alpha)&=G(\alpha)-\beta\ ,&(31)\cr
G_{rs}&=[\beta A^tZA\beta]_{rs}=\beta_r\beta_s\{Z_{ik}-Z_{jk}-Z_{il}+Z_{kl}\}
\ ,&(32)}$$
for $r=(ij)$ and $s=(kl)$.
The level-independence of (29)--(32) can be easily checked using (28). Note
that it is this same quantity $H_{rs}$ that appeared in (9) and (10) [26].

Eqs.~(26), (27), (31) and (32) allow us to see directly why (2) and (11) are
equivalent.
To that end, first notice that the only quantity in the scalar integral in
(11) depending on $p$ is $P(\alpha,p)$. A single operator (15) operating on
$\exp(iP)$ then brings down $(pZA\beta)_r=J_r$, which corresponds to
the replacement $q_r\to J_r$ used in (2).
If $S_0(q)=q_r^\mu q_s^\nu$, then $T(d)=d_r^\mu
d_s^\nu+(i\beta_r/2)\delta_{rs}g^{\mu\nu}$,
so operating on $\exp(iP)$, $T(d)$ brings down
$$J_r^\mu J_s^\nu-{i\over 2}\{[\beta
A^tZA\beta]_{rs}g^{\mu\nu}-\beta_r\delta_{rs}g^{\mu\nu}\}=
J_r^\mu J_s^\nu-{i\over 2}H_{rs}g^{\mu\nu}\ ,\eqno(33)$$
 in agreement with (8) and (9). The equivalence for a more complicated
$S_0(q)$ can be seen similarly.

One might think that only level-independent quantities need be considered in
the Feynman-parameter representation.
Indeed both (2) and (11) contain only level-independent quantities.
This of course does {\it not} mean  that certain calculations cannot be
carried out more easily in one level scheme than another.
In fact,
as we shall see later, the gauge characteristic mentioned in the Introduction
is
borned out only in the particular level scheme described below.

By using the level freedom (25), one can choose $a_i$ so that
$$Z_{ii}=0\quad \forall i\ .\eqno(34)$$
It is in this {\it zero-diagonal level scheme} that the
gauge characteristic emerges.
It is also in this
scheme that
the impedance matrix element $Z_{ij}$ can be computed most easily using
a graphical rule. This rule can be derived from the rule (6) for $P$,
using (27) and (34). If we replace $(\sum p)^2$ in (6) by
$-(\sum p)\.(\sum' p)$, where the two sums are respectively sums of
outgoing momenta attached to the two disjoint trees, then clearly
there are no $p_i^2$ term contributing to (6), and the corresponding
level scheme is therefore given by (34). The graphical rule in the
zero-diagonal scheme is therefore
$$Z_{ij}=-\h\sum_{T_2^{ij}}\prod^{\l+1}\alpha\ ,
\quad(i\not=j)\ ,\eqno(35)$$
where the sum is over the set of all 2-trees $T_2^{ij}$ in which
the vertices $i$ and $j$ lie in two different trees.

The zero-diagonal level scheme of (34) treats all indices symmetrically,
whereas the primitive
level scheme of (24) does not, though the latter has the simplicity of
eliminating an irrelevant degree of freedom in a simple manner.
In these respects the two
level schemes are respectively   analogous to the covariant and physical gauges
in electrodynamics, in that the former    is         manifestly covariant and
the latter contains no longitudinal photons.

We will now derive three {\it level-dependent relations} (eqs.~(43), (47),
and (48)
below), which will be the basis of the string-like formulas of
 Secs.~3 and 4. In this regard it is important to
note that due to external momentum conservation,
neither $J_r$ nor $P$ is a unique function of the $n$ variables $p_i$,
so the derivatives $\p P/\p p_i$ and $\p J_r/\p p_i$, treating each of the $n$
variables $p_i$ as independent, is not uniquely defined. However, as can be
seen in (26) and (27), this
ambiguity is related to the level-dependence of $Z$.
Once a level scheme is defined, such derivatives become meaningful, though
the results are obviously level-dependent. Nevertheless, the
combination of the derivatives in (15) is still level-independent.

We shall call a vertex with {\it two} internal lines
(and any number of external lines) an {\it external
vertex}, and any vertex with more than two internal lines an {\it internal
vertex}. This terminology is unfortunately somewhat misleading because
internal vertices can contain external lines as well.
At an external vertex $a$ we will make the convention that one of the
two internal lines points into the vertex ($A_{ar}=+1$), and the other points
out of the vertex ($A_{ar}=-1$).
The line upstream of the vertex (with
$A_{ar}=+1$) will also be labelled as line $a''$, and the line downstream
of the vertex will be labelled as line $a'$. Current conservation at that
vertex then takes  the form
$$p_a=J_{a''}-J_{a'}\ .\eqno(36)$$
We shall see that external vertices possess special properties
not shared by the internal vertices.

For each external vertex $a$, define the operator
\b{
\p_a&=-\sum_r A_{ar}{\p\over\p \alpha_r}=-{\p\over\p
\alpha_{a''}}+{\p\over\p \alpha_{a'}}\ .&(37)}$$
An important property of an external vertex $a$ is that $\Delta(\alpha),
Z_{ij}$ and $H_{rs}$ depends on $\alpha_{a'}$ only through the combination
$\alpha_{a'}+\alpha_{a''}$, provided $i\not=a\not=j$ and $A_{ar}=A_{as}=0$,
as can be seen from the graphical rules (5), (10), and (35).
As a consequence,
$$\p_a\Delta(\alpha)=\p_aH_{rs}=\p_aZ_{ij}=0\eqno(38)$$
provided $i\not=a\not=j$ and $A_{ar}=A_{as}=0$.

Suppose $a$ is an external vertex and $s$ any internal line. Then
it follows from (10) that
$$H_{a's}=H_{a''s}\ .\eqno(39)$$
To see that
suppose first that lines $a',a'',s$ are all different. Then the claim follows
because there is a 1--1 correspondence between 2-trees $t_2(a's)$ and
$t_2(a''s)$:
instead of cutting the line $a'$ to form a 2-tree, just cut the line $a''$.
Note that lines $a'$ and $a''$ belong to the same loop so they
cannot be cut simultaneously either in $t_2(a's)$ or in $t_2(a''s)$.
 With this 1--1 correspondence,
the equality in (39) follows because $\Delta H_{a's}$ does not contain
$\alpha_{a'}$ and $\Delta H_{a''s}$ does not contain $\alpha_{a''}$.
Now suppose lines $a'$ and $s$ are the same.
Then we must show that $H_{a'a'}=H_{a''a'}$. Since $\Delta H_{a'a'}=-\p \Delta
/ \p \alpha _{a'}$, the terms in $\Delta$ linear in $\alpha_{a'}$ is
$-\alpha_{a'}H_{a'a'}$.
We shall now show that it is also given by $-\alpha_{a'}H_{a''a'}$, thereby
showing that $H_{a'a'}=H_{a''a'}$. To do so,
consider any 2-tree $t_2(a''a')$ used to compute $H_{a''a'}$. Since the
lines $a'$ and $a''$ are now adjacent to each other,
one of the two
trees in $t_2(a''a')$ simply consists of the vertex $a$, and the other is a
tree obtained from the original diagram with the vertex $a$ removed.
There is a 1--1 correspondence between such a tree and a 1-tree of the
original diagram when line $a'$ is cut. Hence the term of $\Delta$
linear in $\alpha_{a'}$ is $-\alpha_{a'}H_{a''a'}$.
This completes the proof of (39).

Eqs.~(38) and (39) are special features of the external vertices not
generally shared by internal vertices. We shall refer to them as
{\it external-vertex relations}.

If $a,b$ are external vertices, then $\p_aP$ can be computed from (29) and
(36) to be
$$\p_aP=-\sum_rA_{ar}J_r^2=-J_{a''}^2+J_{a'}^2=-p_a\.(J_{a''}+J_{a'})\
.\eqno(40)$$
Alternatively, using (28) and (38), one gets
$$\p_aP=\p_a(\sum_{i,j}p_iZ_{ij}p_j)=2p_a\.\p_a(\sum_jZ_{aj}p_j)
\ .\eqno(41)$$
Equating (40) and (41), one concludes that
$$p_a\.[J_{a''}+J_{a'}+2\p_a(\sum_jZ_{aj}p_j)]=0\eqno(42)$$
{\it for every possible} $p$, subject of course only to the restriction
$\sum_ip_i=0$. This means that
$$J_{a''}+J_{a'}=-2\p_a(\sum_jZ_{aj}p_j)=2\p_aV_a
\equiv -2\sum_j\dot Z_{aj}p_j\equiv 2\dot V_a\ .\eqno(43)$$

For an external vertex $a$, it is also useful to define the operator
$$D_a^\mu =\p_a{\p\over\p (p_a)_ \mu }\ .\eqno(44)$$
Then it is clear from (43) that we can also write
$$J_{a''}+J_{a'}=-D_aP\ .\eqno(45)$$

Next, employ (29) and  (30) to compute $\p_a\p_bP$:
$$\p_a\p_bP=2\sum_{r,s}A_{ar}A_{bs}J_rH_{rs}J_s\ .\eqno(46)$$
Since $a,b$ are external vertices, (40) is true, so using  (36) and (39),
we get
 $$\p_a\p_bP=2p_a\.p_bH_{a'b'}\ .$$
Suppose $a\not=b$. Then an alternative calculation is to use (38):
$$\p_a\p_bP=\p_a\p_b(\sum_{i,j}p_iZ_{ij}p_j)
=2p_a\.p_b\p_a\p_b(Z_{ab})\ .$$
Again these two expressions must be equal for all momentum configurations,
so
$$H_{a'b'}g^{\mu \nu }=\p_a\p_b(Z_{ab})g^{\mu \nu }=
\h D_a^\mu D_b^\nu P\equiv \ddot Z_{ab}g^{\mu \nu }\quad (a\not=b)
\ ,\eqno(47)$$
where (38) has again be used to obtain the last equality.
For $a=b$, (47) is not longer valid. Instead, (39) can be used to
relate $H_{a'a'}$ to $H_{a'a''}$, which can then be calculated using (47).

There is one more relation for an external vertex $a$ which we need,
$$D_a^\mu J_s^\nu =-H_{rs}g^{\mu \nu }\ ,\quad (A_{as}=0)\ ,\eqno(48)$$
which is true for $r=a'$ or $r=a''$, and for $a'\not=s\not=a''$.
To prove it, consider a situation where the primitive spin factor is
$S_0(J)=-(J_{a'}+J_{a''})^\mu J^\nu _s$. The modified spin factor $S(J)$
is, according to (9) and (39), given by $S(J)=S_0(J)+iH_{rs}g^{\mu \nu }$.
On the other hand, using (11)--(15), and the assumption that $a'\not=
s\not=a''$, $S(J)$ is also given by
$\exp(-iP)S_0(d)\exp(iP)$. Using (45), this becomes
$-i\exp(-iP)d_s^\nu D_a^\mu \exp(iP)= -i\exp(-iP)D_a^\mu J_s^\nu \exp(iP)$,
which is equal to
$-i\(D_a^\mu J_s^\nu \)-(J_{a'}+J_{a''})^\mu J^\nu _s$.
Comparing these two ways of obtaining $S(J)$, (48) follows.

As it stands, (48) is not true when $s=a'$ or $a''$.
For example, using the explicit example to be considered in Sec.~5,
one can show explicitly that when $a=1$ and $s=2$, only half of the
rhs of (48) is obtained from the lhs.

It is this lack of universal validity of (48) that
restricts the validity of the string-like formula, as we shall see in
the next section.

Because of the importance of this restriction it should also be pointed out
that there is a false derivation of (48)
which makes it appear to be valid for {\it all} $a$ and $s$.
The false argument follows from (30), (36), (37), (39), and (44).
It is false because strangely enough, (36) is not a valid equation
for the present application, where $D_a$ and hence derivative of
external momenta are involved. Eq.~(36) is of course valid in
the physical case when
external momentum conservation (28)  is used. On the other
hand, since the operation involving $D_a$ is level-dependent, we must
not use this conservation law before the momentum differentiation.
In that case the rhs of (36) could be equal to, say, $-\sum_{b\not=a}p_b$,
which is not at all the same thing as the lhs of (36) as far as momentum
differentiation is concerned. The correct way to check the validity
of (36) is to compute the rhs of (36) using (26), sticking to whatever
level scheme one is using and refraining from ever using external
momentum conservation before the momentum differentiation.
Without using momentum conservation, the currents $J$ is no longer
level-independent, hence the outcome of the rhs of (36) depends
on the level scheme one uses. In this way one can show by explicit
examples that (36) is generally invalid in the zero-diagonal level scheme.
 This observation once
again shows the subtlety of the level dependence problem.

Let us turn to
amplitudes in the configuration space. Since field
theories are local in $x$, interactions should
look simpler  in the $x$-space than in the $p$-space.
For conceptual reasons
it is therefore worthwhile to look at the expressions in the $x$-space,
although in practical calculations we must return to the momentum space.
Fourier-transforming the momentum-space amplitude (1), we get
\b{
(2\pi)^4
i^lM(p)\delta^4\(\sum_{i=1}^np_i\)&=\int\(\prod_{i=1}^nd^4x_i\)
\exp\[i\sum_{i=1}^np_i\.x_i\]M'(x)\ ,&(49)\cr
M'(x)&=S_0(-i\p/\p y)\prod_{r=1}^N\Delta_+(y_r)\ ,&(50)}$$
where the scalar propagator is given by
$$\eqalignno{
\Delta_+(y_r)&={1\over (2\pi)^4}\int d^4q_r{\exp[iq_r\. y_r]\over
m_r^2-q_r^2-i\epsilon}&\cr
&={1\over 16\pi^2}\int_0^\infty d\beta_r\exp\[-i{m_r^2\over\beta_r}
+{i\over 4}\beta_ry_r^2\]&\cr
&={1\over 16 \pi ^2}\int_0^\infty {d \alpha _r\over \alpha _r^2}
\exp\[-i \alpha _rm_r^2+i{y_r^2\over 4 \alpha _r}\]\ .&(51)}$$
Hence
\b{
M'(x)&=\({1\over 16\pi^2}\)^N\int_0^\infty\(\prod_{r=1}^Nd\beta_r\)
S_0\(\h \beta_ry_r\)\exp\[-i\sum_{r}\alpha_rm_r^2+iP\]\ ,&\cr
P&={1\over 4}\sum_{r=1}^N\beta_ry_r^2\ .&(52)}$$
For a concrete example see eq.~(70) below.
In (50) and (52), we must substitute $y_r=x_i-x_j$ for an $r=(ij)$
and as usual, $\beta_r=1/\alpha_r$. In the
electric
circuit analogy, $x_i/2$ is really the potential $V_i$ at vertex $i$,
so $y_r/2$ is
the potential drop $v_r$ across resistance $r$ [26, 29--32]. $P$ is again
equal to the power consumed by the network.
One can also obtain similar
formulas and interpretations when some of the $x_i$'s are integrated over.
For details, see [26].

A Feynman diagram can also be given a different interpretation
[26, 29--32] as a maze in which a particle moves in. The
off-shell four-momentum of the particle along an internal line  $r$ is
$J_r^\mu$; the four-distance it travels along $r$ is
$y_r^\mu =x_i^\mu -x_j^\mu $, and this is accomplished in an amount of
`proper time' equal to $2m\alpha_r$.
Incidentally, it is perhaps more appropriate  to think of the
object travelling around the maze as a quantum
mechanical wave rather than a classical particle because it is off-shell,
and because it is easier to think of a wave splitting  and recombining
at vertex junctions than a classical particle.
In this interpretation, the integrand in the last expression of (51)
is essentially
the overlap matrix element $\bk{x_i \tau _i|x_j \tau _j}$
under the Hamiltonian $m(\dot x(\tau)^2+1)/2$ where $\tau =
\tau _i-\tau _j=2m \alpha _r$ is the proper time elaspsed.

This interpretation exists for all diagrams and it can perhaps
be thought of as the basis for a multiloop generalization of the string
approach [22--24] and the first quantized approach [25].
There is however a fundamental difference
between a string propagation on a multi-genus worldsheet and
this particle propagating in the Feynman-diagram maze.
The former satisfies the wave equation everywhere on the worldsheet,
whereas the latter can be described by free-particle equation of motion
only between vertices. Vertices are singularities where
interactions take place, where free-particle equation of motion breaks down.
No such singularities are present on the
worldsheet on account of string's conformal invariance [33], but this
conformal invariance is lost in the infinite tension limit when  the
worldsheet collapses into a network of
worldlines because the $\sigma $-variable
along the string disappears.
Reparametrization invariance in the $\tau $ variable
can however still be kept as is done in [25], thereby preserving
the string characters and the string-like formulas.
As a result of the difference, although a string amplitude can
be written as a path integral of a
free-string over  multi-genus worldsheets, a free-particle path
integral no longer exists for Feynman diagrams
with internal vertices, unless their associated singularities and
interactions can somehow be incorporated, as is done in the
Feynman-parameter representation.

\bigskip

\centerline{\bf 3. Scalar QED}

There are two kinds of electromagnetic
vertices in scalar electrodynamics: the cubic vertex of Fig.~1,
$$C_j=e \epsilon_j\.(q_{j''}+q_{j'})\ ,\eqno(53)$$
and the quartic (seagull) vertex of Fig.~2,
$$Q_j=e^2 \epsilon_j\. \epsilon_{j+1}\ ,\eqno(54)$$
with $\epsilon_j$ being the
polarization vector for the $j$th photon.
We are interested in finding a string-like formula for {\it standard diagrams},
which by definition are diagrams without seagull vertices.

Fig.~3 represents a one-loop $n$-photon standard diagram.
All other one-loop standard diagrams are  obtained from it  by
permuting the external photon lines.
The amplitude for Fig.~3 can be derived from the string or
the first quantized
approach [23,25] to be
\b{
M(p)=&-{e^n\over 16\pi^2}
\int (\prod_{i=1}^ndt_i)  t_n^{-2}&\cr
&\bk{\exp\left\{ -im^2t_n+i\sum_{i<j}\[p_i\.p_jG_B^{ij}-
(\epsilon_i\.p_j
-\epsilon _j\.p_i)\dot G_B^{ij}+\epsilon_i\.\epsilon_j\ddot G_B^{ij}\]\right\}
}\ . &(55)}$$
This formula resembles
a formula first obtained by Bern and Kosower [22] from string theory
for QCD so it shall be referred to  as a {\it string-like formula}.
The integration region in (55) is
$$0\le t_1\le t_2\le\cdots\le t_n<\infty\ ;\eqno(56)$$
the notation $\bk{f(\epsilon_1,\epsilon_2,\cdots,\epsilon_n)}$ means that
 only terms multilinear in all the $\epsilon_i$'s in $f$ should be kept.
In other words, if $\theta_i$ are Grassmann variables,
$$\bk{f(\epsilon_1,\epsilon_2,\cdots,\epsilon_n)}=\pm\int d\theta_1 d\theta_2
\cdots
d\theta_nf(\theta_1\epsilon_1,\theta_2\epsilon_2,\cdots,\theta_n\epsilon_n)\ .
\eqno(57)$$
The function $G_B^{ij}$ and its derivatives are  functions of
$t_{ij}=t_i-t_j$ whose explicit expressions for $i<j$ are
\b{
G_B^{ij}&=t_{ij}(t_n+t_{ij})/t_n\ ,&(58)\cr
\dot G_B^{ij}&={\p G_B^{ij}\over\p t_i}=(t_n+2t_{ij})/t_n\ ,&(59)\cr
\ddot G_B^{ij}&={\p^2G_B^{ij}\over\p t_i^2}=2/t_n\ .&(60)}$$
One can also add in $\delta $-functions to simulate the seagull
vertex contributions [25] but we shall not do that here.

What is crucial in this formula is that  the various
terms appearing in the exponent between square
brackets are all given by the
functions $G_B^{ij}$ and their derivatives. It is this {\it gauge
characteristic}
that allows the integration by parts (IBP) technique discussed in the
Introduction to be carried out.
We shall come
back to a discussion of the gauge characteristic after we derive the
 multiloop string-like formula
in scalar QED that exhibits this feature.

As we shall see below, a string-like formula with the gauge characteristic
exists for any process to any
number of loops, provided the following technical restriction
 is obeyed.
The formula turns out to look very similar to (55) when no derivative
couplings are present, {\it i.e.}, when there are no internal
photon lines and when the `strong  interaction' between scalar
particles contains no derivative couplings. In the more general
situation, a string-like formula with the gauge characteristic still exists,
but it looks a bit more complicated.

Derivative couplings are present in electromagnetic
cubic vertices (53), and perhaps in the `strong interactions'
between the scalar particles. All of them contribute to the
primitive spin factor $S_0(q)$ in eq.~(1). The technical restriction
mentioned above
is imposed so that eq.~(48) can be used. As we shall see below,
this means that if $a$ is an {\it external}
vertex, then $S_0(q)$ should {\it not} contain any $q_{a'}$ or
$q_{a''}$ other than those in $C_a$ of (53). The notations are
those used in Sec.~2:
$a''$ and $a'$ are respectively the lines pointing into and out of
the external vertex $a$. This restriction means
for example that if there are
internal photon propagators, then these internal vertices should not
be adjacent to an external vertex.

To obtain a string-like formula for multiloops, it is
necessary to find out first what is the generalization of $G_B^{ij}$. By
comparing
the terms quadratic in $p$ in the exponents of (55) with (2) and (27),
it is
clear that if anything works $G_B^{ij}$ has to be $Z_{ij}/2$.
However, this alone does not tell us in what level-scheme should $Z_{ij}$
be expressed  so as to capture the gauge characteristic. Before we solve
this problem there is however a second problem that needs to be
considered.

In one loop, $G_B^{ij}$ is a function of the `time' difference
$t_{ij}=t_i-t_j$. Time is connected to the Feynman parameters
of Fig.~3 by
$$t_i=\sum_{k=1}^i\alpha_k\ .\eqno(61)$$
For multiloop diagrams, $Z_{ij}$ is a rather complicated function of
$\alpha_i$, and it is impossible in general to define  a `time' parameter
$t_i$ so that $Z_{ij}$ is a function only of $t_i-t_j$. This is related
to the fact that the particle in the more general diagram has to split
and recombine. Given that,
the next important
question to solve for multiloops is to determine what should replace the
time derivatives in (59) and (60).
It turns out that the time derivative
$\p/\p t_a$ should in the general case be replaced by $\p_a$ defined in (37),
and in so doing the gauge characteristic will be preserved {\it provided}
that $Z_{aj}$ is expressed in the zero-diagonal level scheme of (34).

We shall use the indices $a,b$ to denote cubic electromagnetic
vertices (53) with an external photon line, and the indices $i,j$
to label {\it all} the vertices. Note that $a,b$ are external vertices
in the sense of Sec.~2, {\it viz.,} two internal lines are connected to
each of them.
 In the simpler case when
internal photon lines and derivative couplings between the scalar
particles are absent,
the multiloop string-like formula for any multiloop standard diagram is
$$\eqalignno{
M(p)=&{i^{N-n}e^n\over (-16\pi^2)^\l}
\int_0^\infty (\prod_{r=1}^Nd\alpha_r)  \Delta(\alpha )^{-2}S''_0&\cr
&\bk{\exp\left\{-i\sum_{r=1}^N \alpha _rm_r^2+i\sum_{i,j}p_i\.p_jZ_{ij}
-2\sum_{a,j}
\epsilon _a\.p_j\dot Z_{aj}+
\sum_{a,b}\epsilon _a\.\epsilon _b\ddot Z_{ab}\right\}
}\ ,&(62)}$$
where
$$\dot Z_{aj}=\p_aZ_{aj}\ ,\quad \ddot Z_{ab}=\p_a\p_bZ_{ab}\ ,\eqno(63)$$
with $Z_{ii}=0$ and $\p_a$  defined in (37). A similar but slightly
more complicated formula ((76) below) exists when derivative couplings
are present, as long as the technical restriction mentioned earlier is
obeyed.

We will proceed now to demonstrate (62).
The primitive spin factor $S_0(J)$ for any standard diagram is now given by
$S_0(J)=S'_0(J)S''_0$, where
$$S'_0(J)=\prod_aC_a=\prod_{a}^n[e \epsilon _a\.(J_{a'}+J_{a''})]=
(-ie)^n\bk{\exp[i\sum_{a}\epsilon_a\.(J_{a'}+J_{a''})]}\ \eqno(64)$$
is the primitive spin factor for the external cubic electromagnetic
vertices, and $S''_0$ is the momentum-independent vertex factors
of the rest.
Using (43), this becomes
$$S'_0(J)=
(-ie)^n\bk{\exp[-2i\sum_{a,j}\epsilon_a\.p_j\dot Z_{aj}]}\ .\eqno(65)$$
To compute the modified spin factor $S(J)$, the contraction rule (9)
is first used to compute the contraction of a pair of electromagnetic vertices.
After using (47), this becomes
$$C_aC_b\to
-2ie^2\epsilon_a\.\epsilon_bH_{ab}=-2ie^2\epsilon_a\.\epsilon_b\ddot Z_{ab}\ .
\eqno(66)$$
Then (8) is used to sum up all the contractions, giving
$$S(J)=(-ie)^nS''_0\bk{\exp\{i\sum_{a,j} [-2\epsilon_a\.p_j\dot
Z_{aj}]+i\sum_{a,b}[\epsilon_a\.\epsilon_b\ddot Z_{ab}]}\}\ .\eqno(67)$$
Substituting this into (2), we obtain the string-like formula (62).

The string formula (62) can be simplified by noting that
$$\eqalignno{
S(J)\exp(iP)&=(-ie)^n\bk{\exp\[-\sum_{a}\epsilon_a\.D_a\]}S''_0
\exp(iP)\ ,&(68)\cr
P&=\sum_{i,j}p_iZ_{ij}p_j\ ,&(69)}$$
where $D_a$ is defined in (44).

{}From the discussions at the end of Sec.~2, one expects the amplitude
to be particularly simple when expressed
in the configuration space.
Indeed,  using  (69) and the definition of the configuration-space amplitude
(49), one gets
$$M'(x)=(-ie)^n\bk{\exp[-i\sum_a\epsilon_a\.\p_ax_a]}
\prod_{r=1}^N\Delta_+(y_r)\ .
\eqno(70)$$

The string-like formula (62)
 exhibits a simple gauge transformation property, which
is probably the reason why it is inherently important.
This is the {\it gauge characteristic} we have been talking about.
This gauge property
is particularly transparent when (62) is written in the form of (68),
for then a gauge transformation
$$\epsilon _a\to \epsilon _a+p_a \lambda _a\eqno(71)$$
simply brings on the following change of the integrand of (2):
\b{ &(-ie)^n
\bk{\exp[-\sum_a\epsilon_a\. D_a]}\Delta ^{-2}S''_0
\exp(-iD)\to&(72)\cr &(-ie)^n
\bk{\exp[-\sum_a(\epsilon_a+\lambda _ap_a)\.D_a]}\Delta ^{-2}
S''_0\exp(-iD)\
.&(73)}$$
Note that the first equation in (38) has been used, and we have also used
the fact that lines $a'$ and $a''$ have the same mass.
Since the gauge parameters $\lambda _a$ are multiplied by
the derivatives $\p_a=\p/\p \alpha _{a'}-\p/\p \alpha _{a''}$, they appear only
in surface terms corresponding to diagrams with $\alpha _{a'}$ or
$\alpha _{a''}$ short-circuited. It is these surface terms that
connect the permuted standard
diagrams and the seagull diagrams to enable gauge-dependent terms to be
cancelled out at the end. In this way the string-like formula
 makes the Ward-Takahashi  identity in Feynman-parameter
space almost explicit as the one seen in momentum space, and it is also
this same gauge characteristic which enables the IBP technique to be
applied.

We turn now to the  general case
where internal photon lines and
 `strong interaction' with derivative couplings are allowed to be present
in the standard diagram, provided the technical restriction mentioned
earlier is met.
The primitive spin factor is then given by
$$S_0(J)=S'_0(J)S''_0(J)\ ,\eqno(74)$$
where $S'_0(J)$ is the contribution from the external electromagnetic vertices
which is given by (64),
and $S''_0(J)$ is the contribution from the rest of the vertex
factors which is now momentum dependent. The modified spin factor
$S(J)$ is obtained from (74) by binary contractions with the rule of
(9). When the technical restriction is met,
 contractions with the currents at an external vertex $a$
may be accomplished by the operator $D_a$ as in (48),
hence we obtain a relation similar
to (68), which now reads
$$S(J)\exp(iP)=(-ie)^n\bk{\exp\[-\sum_a\epsilon_a\.D_a\]}S''(J)\exp(iP)\
,\eqno(75)$$
where  $S''(J)$ is the modified spin factor corresponding to the primitive spin
factor $S''_0(J)$ of (74). Substituting this into (2), we finally obtain
the string-like formula in the general case to be
$$\eqalignno{
M(p)=&{i^{N-n}e^n\over (-16\pi^2)^\l}
\int_0^\infty (\prod_{r=1}^Nd\alpha_r)
\bk{\exp\[-\sum_a\epsilon_a\.D_a\]}\ \.&\cr
& \Delta(\alpha )^{-2}S''(J)\exp\[-i\sum_{r=1}^N\alpha_rm_r^2
+i\sum_{i,j}p_i\.p_jZ_{ij}\]\ .&\cr&&(76)}$$
Again, the first equality of (38) as well as the fact electromagnetic
interaction is diagonal in mass have been used.
The gauge transformation property of this is similar to (73). Note that
$\exp[-\sum_a\epsilon_a\.D_a]$ as defined in (44) is a translation
operator shifting momentum $p_a$ by an amount $-\epsilon_a\.\p_a$.
If we carry out this momentum shift in (76), the exponential will
return to the form (62), but the momenta $p$ implicitly contained
in $S''(J)$ must be so shifted as well, and this shift contains the
gauge-dependent quantity $\epsilon_a$. We prefer not to write
it in this shifted form for it makes  the gauge
tranformation property more obscure.

\bigskip
\centerline{\bf 4. Spinor QED}

In spinor electrodynamics, an $n$-photon one-loop amplitude has a string-like
formula given in [23,25]. To obtain its multiloop generalization, a Gordon
decompostion has to be made to separate the current into a
convective part and a spin part.

Consider a fermion propagating in the presence of a background electromagnetic
potential $A^\mu(x)$ and a background neutral scalar field $\phi(x)$.
Depending on what is required, we can later on replace
the beackground $A^\mu(x)$ by a polarization vector or one end of an internal
photon propagator, and the background $\phi(x)$ by an external or internal
`neutral scalar meson' coupled to the fermion by `strong interaction'.
For definiteness we shall assume a Yukawa coupling for the strong interaction,
but this point is not crucial for the following discussion.
The fermion propagator is  $[m-i\gamma(\p-ieA)+\phi]^{-1}\delta^4(x-y)$;
perturbation series are obtained by expanding this in power series of $A$
and $\phi $.
Gordon decomposition is accomplished by noting that
\b{
[m+\phi-i\gamma(\p-ieA)]^{-1}&=[m+\phi+i\gamma(\p-ieA)]\.I\ ,&(77)}$$
where
\b{
I&=
\{[m+\phi-i\gamma(\p-ieA)][m+\phi+i\gamma(\p-ieA)]\}^{-1}&\cr
&=\{(m+\phi)^2-i\gamma(\p\phi)+(\p-ieA)^2-{e\over
2}\sigma^{\mu\nu}F_{\mu\nu}\}^{-
1 } & \cr
&=(m^2+\p^2)^{-1}\sum_{n=0}^\infty[(C+Q+S+M_1+M_2+M_3)(m^2+\p^2)^{-1}]^n
\ ,&(78)}$$
and
\b{
\sigma^{\mu\nu}&={i\over 2}[\gamma^\mu,\gamma^\nu]\ ,&\cr
C&=ie(\p A+A\p)&\cr
Q&=e^2A^2&\cr
S&={e\over 2}\sigma^{\mu\nu}F_{\mu\nu}&\cr
M_1&=-2m\phi&\cr
M_2&=-\phi^2&\cr
M_3&=i\gamma(\p\phi)
\ .&(79)}$$
$C$ and $Q$ are just the cubic and the seagull electromagnetic vertices for a
scalar particle. They give rise to the convective part of the current.
On top of these, there is
the spin vertex $S$ which gives rise to the magnetic
moment of the fermion.  The vertices $M_i$ are strong interaction vertices
between the fermion and the
neutral scalar meson. Note that $M_3$ is both momentum and spin dependent.

In a scattering diagram,
a fermion propagator can end in two  external fermions,
in which case the relevant factor  is
\b{
&\bar u(m-i\gamma\p)[m-i\gamma(\p-ieA)+\phi]^{-1}(m-i\gamma\p)u
=&\cr
&{1\over 2m}\bar
u\sum_{n=0}^\infty[(C+Q+S+M_1+M_2+M_3)(m^2+\p^2)^{-1}]^n(m^2+\p^2)u\ ,&(80)}$$
or with the wave functions $u,\bar u$ replaced by $v,\bar v$. Otherwise,
it can close on itself in a loop, in which case the proper factor is
\b{
\log\{\det[m-i\gamma(\p-ieA)+\phi]\}&=\h\log\{\det[I]\}&\cr
&=\h\Tr\{\log\[1+\sum_{n=1}^\infty
\((C+Q+S+M_1+M_2+M_3)(m^2+\p^2)^{-1}\)^n\]\}\ .&\cr
&&(81)}$$
A constant irrelevant normalization factor has been dropped in
the last expression.
In other words, other than signs associated with statistics,
the electromagnetic interaction of
spinor QED differs from that of scalar QED only in having the
extra vertex $S$.

The rest of the discussion is identical to the scalar case, leading up
to the string-like equation (76) for standard diagrams, where $S''(J)$ is now
the modified spin factor for the vertices $S$, $M_i$, and the internal
$C$'s. The technical restriction mentioned in Sec.~3 is automatically
fulfilled for the derivatively-coupled vertex $M_3$ because it depends only
on the momentum of the neutral meson but not of the fermions.

To establish contact with the formula in [22--25], let us specialize to the
 a
one-fermion-loop $n$-photon amplitude in the absence of all strong
interactions. Then
$$S''(J)=\prod_cS_c=\tr\bbl\prod_c\[-ie\sigma_{\mu\nu}\epsilon_c^\mu
p_c^\nu\]\bbr\ .\eqno(82)$$
The trace of a product of $2m$ $\gamma$-matrices is given by 4 times the
sum of all signed contractions, with each signed contraction
 given by a product of $m$ factors of $(\pm g_{\alpha\beta})$'s,
corresponding to
 the contraction of $(\cdots\gamma_\alpha\cdots\gamma_\beta\cdots)$.
Using this rule, it is easy to compute the trace
$${1\over 4}\({2\over i}\)^m\tr\[(\epsilon_1\.\sigma\.p_1)
\cdots(\epsilon_m\.\sigma\.p_m)\]\eqno(83)$$
as follows. Take any permutation $t\in P_m$ of $m$ objects and express it into
cycles: $t=(t_1t_2\cdots t_k)(\cdots)\cdots$. Then (83)
is given by
$$\eta_t\[(\epsilon_{t_1}\.p_{t_2})(\epsilon_{t_2}\.p_{t_3})(\cdots
)(\epsilon_{t_k}\.p_{t_1})\]\[\cdots\]\cdots\eqno(84)$$
summed over all permutations $t\in P_m$, and summed over all possible
interchanges between every pair
$$\epsilon_{t_i}\leftrightarrow p_{t_i}\ .\eqno(85)$$
$\eta_t$ is the signature of the permutation, being $\pm 1$ for even/odd
permutations, and a minus sign is to be associated
with each interchange (85).

It is these terms in (85) that give rise to the functions $G_F^{ij}$
and their associated rules in the formula of [22--25].

\bigskip
\centerline{\bf 5. An Example}

To illustrate some of the circuit quantities and relations,
let us consider the two-loop Compton amplitude of Fig.~4.
The dash lines are photons, and the solid lines are
charged scalar mesons. The external vertices are $a=1,3$,
and the `internal' vertices are $j=2,4$. Using (35), one
obtains the impedance matrix in the zero-diagonal scheme
to be
\b{
\Delta&=(\alpha_1+\alpha_2)(\alpha_3+\alpha_4)+\alpha_5(\alpha_1+
\alpha_2+\alpha_3+\alpha_4)&\cr
Z_{ii}&=0&\cr
Z_{ij}&=Z_{ji}&\cr
-2\Delta Z_{12}&=\alpha_2 [(\alpha_3 + \alpha_4)(\alpha _1+\alpha _5) +
 +\alpha _1 \alpha_5]&\cr
-2\Delta Z_{13}&= (\alpha_1 \alpha_2 \alpha_3 + \alpha_1 \alpha_2
\alpha_4 + \alpha_1 \alpha_3 \alpha_4 + \alpha_2 \alpha_3 \alpha_4) +
\alpha_5(\alpha_1+\alpha_4)(\alpha_2+\alpha_3) &\cr
-2\Delta Z_{14}&=
\alpha_1 [(\alpha_3 +  \alpha_4)(\alpha _2 +  \alpha_5) +
\alpha _2\alpha_5]&\cr
-2\Delta Z_{23}&=
\alpha_3[(\alpha_1  + \alpha_2 )(\alpha _4+  \alpha_5) +
\alpha_4\alpha_5 ]&\cr
-2\Delta Z_{24}&=
\alpha_5(\alpha_1 + \alpha_2) (\alpha_3 + \alpha_4)&\cr
-2\Delta Z_{34}&=
\alpha_4[(\alpha_1  + \alpha_2)(\alpha _3 + \alpha_5) +
\alpha_3  \alpha _5)]\ .&(86)}$$
{}From these and the definition (37), one can compute the various
derivatives $\dot Z_{aj}=\p_aZ_{aj}$ and $\ddot Z_{ab}=\p_a\p_b Z_{ab}$
to be
\b{
-2\Delta\dot Z_{12}&=(a_2-\alpha _1)(\alpha _4+\alpha _4)+
\alpha _5(-\alpha _1+\alpha _2-\alpha_3-\alpha_4)&\cr
-2\Delta\dot Z_{13}&=(\alpha_2-\alpha_1)(\alpha_3+\alpha_4)
+\alpha_5(-\alpha _1+\alpha_2+\alpha_3-\alpha_4)&\cr
-2\Delta\dot Z_{14}&=(\alpha_2-\alpha_1)(\alpha_3+\alpha_4)+\alpha_5
(-\alpha _1+\alpha _2+\alpha _3+\alpha _4)&\cr
-2\Delta\dot Z_{31}&=(\alpha_4-\alpha_3)(\alpha_1+\alpha_2)
+\alpha_5(\alpha_1-\alpha_2-\alpha_3+\alpha_4)&\cr
-2\Delta\dot
Z_{32}&=(\alpha_4-\alpha_3)(\alpha_1+\alpha_2)+\alpha_5(
\alpha _1+\alpha _2-\alpha _3+\alpha _4) &\cr
-2\Delta\dot
Z_{34}&=(\alpha_4-\alpha_3)(\alpha_1+\alpha_2)+\alpha_5(
-\alpha _1-\alpha_2-\alpha _3+\alpha _4) &\cr
-2 \Delta\ddot Z_{13}&=2 \alpha _5\ .&(87)\cr}$$
Note that the one loop $n$-photon relationship $\dot Z_{13}=-\dot Z_{31}$
is no longer valid.

These are all the quantities needed in the string-like formula
(62). The string-like formula is obtained by
using the external-vertex formulas (38) and (39), as well as
the level-dependent relations (43), (47), and (48).
Let us look into the explict form of these relations for

the external vertex $a=1$
in the present example. For $a=1$, we see from Fig.~4 that
$a'=1$ and $a''=2$, so $\p_1=\p/\p \alpha_1-\p/\p \alpha_2$.
The quantities relevant to the level-dependent relations are
\b{
\Delta(J_1+J_2)=2\dot V_1&=(\alpha _1-\alpha _2)
(\alpha _3+\alpha _4+\alpha _5)p_1&\cr
&+\alpha _5[(\alpha _3+\alpha _4)(p_4-p_2)+(\alpha _3-\alpha _4)p_3]&\cr
\Delta ( D_1^\mu J_1^\nu )=\Delta (D_1^\mu J_2^\nu )
&=\h(\alpha _3+\alpha _4+\alpha _5)g^{\mu \nu }
=-\h \Delta H_{11}g^{\mu \nu }=-\h\Delta H_{12}g^{\mu \nu }&\cr
\Delta (D_1^\mu J_3^\nu )=\Delta (D_1^\mu J_4^\nu )
&=\alpha _5g^{\mu \nu }=-\Delta H_{13}g^{\mu \nu }=
-\Delta H_{14}g^{\mu \nu }=-\Delta
\ddot Z_{13}g^{\mu \nu }&\cr
\Delta (D_1^\mu J_5^\nu )&=(\alpha _3+\alpha _4)g^{\mu \nu }
=-\Delta H_{15}g^{\mu \nu }\ .&(88)}$$
Consequently (43) and (47) are valid, as they should be, and (48) is true
when $a'\not=s\not=a''$, but is not true when $s=a'$ or $a''$. In the
latter cases an extra factor of 1/2 appears. From (86) and (88),
it is also easy to check explicitly the external-vertex
relations (38) to be correct. We have not written down $H_{2s}$, but
they can be calculated explicitly to see that (39) is indeed valid.

\bigskip
\centerline{\bf Acknowledgement}
I am grateful to Zvi Bern, Matt Strassler, Frank Wilczek,
and Tung-Mow Yan for stimulating
discussions. This research is supported in part by the Natural Sciences and
Engineering Research Council of Canada and the Qu\'ebec Department of
Education.

\vfill\eject
\centerline{\bf References}
\bigskip
\def\i#1{\item{[#1]}}
\def\npb#1{{\it Nucl.~Phys. }{\bf B#1}}
\def\plb#1{{\it Phys.~Lett. }{\bf #1B}}
\def\prl#1{{\it Phys.~Rev.~Lett. }{\bf B#1}}
\def\prd#1{{\it Phys.~Rev. }{\bf D#1}}
\def\pr#1{{\it Phys.~Rep. }{\bf #1}}
\def\ibid#1{{\it ibid.} {\bf #1}}
\parskip0pt
\i 1 F.A. Berends, R. Kleiss, P. De Causmaecker, R. Gastmans,
W. Troost, and T.T. Wu, \plb{103} (1981), 124;
\npb{206} (1982), 61;
\ibid{239} (1984), 382; \ibid{239} (1984), 395; \ibid{264} (1986), 243;
\ibid{264} (1986), 265.
\i 2  P. De Causmaecker, R. Gastmans,
W. Troost, and T.T. Wu, \plb{105} (1981), 215; \npb{206} (1982), 53.
\i{3} Z. Xu, D.-H. Zhang, and L. Chang, Tsinghua University Preprints,
Beijing, China, TUTP-84/4, TUTP-84/5, TUTP-84/6; \npb{291} (1987), 392.
\i 4 F.A. Berends and W.T. Giele, \npb{294} (1987), 700; \ibid{306} (1988),
759; \ibid{313} (1989), 595.
\i 5 F.A. Berends, W.T. Giele, and H. Kuijf, \plb{211} (1988), 91; \ibid{232}
(1989), 266; \npb{321}
(1989), 39;\ibid{333} (1990), 120.
\i 6 J. Gunion and J. Kalinowski, \prd{34} (1986), 2119.
\i {7} J. Gunion and Z. Kunszt, \plb{159} (1985), 167; \ibid{161} (1985), 333;
\ibid{176} (1986) 477.
\i{8} K. Hagiwara and D. Zeppenfeld, \npb{313} (1989), 560.
\i{9} R. Kleiss and H. Kuijf, \npb{312} (1989), 616.
\i{10} R. Kleiss and W.J. Stirling, \npb{262} (1985), 235.
\i{11} D. Kosower, \npb{315} (1989), 391; \ibid{335} (1990), 23; \plb{254}
(1991) 439.
\i{12} J.G. K\"orner and P. Sieben, Mainz preprint MZ-TH/90-08.
\i{13} Z. Kunzt, \npb{271} (1986), 333.
\i{14} M. Mangano, \npb{309} (1988), 461.
\i{15} M. Mangano, S. Parke, and Z. Xu, \npb{298} (1988), 653.
\i{16} M. Mangano and S.J. Parke, \npb{299} (1988), 673; \prd{39} (1989), 758;
       \pr{200} (1991), 301.
\i{17} Z. Bern and D.K. Kosower, \npb{362} (1991), 289.
\i{18} S. Parke and T. Taylor, \plb{157} (1985), 81; \npb{269} (1986), 410;
\prl{56} (1986), 2459; \prd{35} (1987), 313.
\i{19} C. Dunn and T.-M. Yan, \npb{352} (1989), 402.
\i{20} G. Mahlon and T.-M. Yan, Cornell preprints CLNS91/1119; 91/1120.
\i{21} G. Mahlon, Cornell preprint CLNS 92/1154; Cornell University Ph.D.
thesis.
\i{22} Z. Bern and D.K. Kosower, \prl{66} (1991), 1669; \npb{379} (1992) 451;
 preprint Fermilab-Conf-91/71-T.
\i{23} Z. Bern and D.C. Dunbar, \npb{379} (1992) 562.
\i{24} Z. Bern, L. Dixon, and D.A. Kosower, preprint SLAC-PUB-6012
(UCLA/92/TEP/47, CERN-TH.6733/92).
\i{25} M. Strassler, \npb{385} (1992) 145; SLAC
preprint (in preparation).
\i{26} C.S. Lam and J.P. Lebrun, {\it Nuovo Cimento} {\bf 59A} (1969), 397.
\i{27} C.S. Lam, preprint McGill/92-32 (hep-ph/9207266).
\i{28} J. Paton and Chan Hong-Mo, \npb{10} (1969), 519.
\i{29} J.D. Bjorken, Standford Ph.D. thesis (1958).
\i{30} J.D. Bjorken and S.D. Drell, `Relativistic Quantum Fields'
(McGraw-Hill, 1965).
\i{31} J. Mathews, {\it Phys.~Rev.} {\bf 113} (1959) 381.
\i{32} S. Coleman and R. Norton, {\it Nuovo Cimento} {\bf 38} (1965) 438.
\i{33} M.B. Green, J.H. Schwarz, and E. Witten, `Superstring Theory'
(Cambridge University Press, 1987).
\vfill\eject
\centerline{\bf Figure Captions}
\bigskip
\item{Fig.~1:} The cubic electromagnetic vertex in scalar QED.
\bigskip
\item{Fig.~2:} The seagull electromagnetic vertex in scalar QED.
\bigskip
\item{Fig.~3:} A one-loop $n$-photon amplitudes with outgoing external
momenta $p_i$. $n=6$ is shown in the diagram. The numbers around
the loop label the internal lines.
\bigskip
\item{Fig.~4:} A two-loop Compton scattering amplitude in scalar
QED. The $p_i$ are the external outgoing momenta, and the five
internal lines are numbered as shown.

\end